\documentclass[12pt]{iopart}
\usepackage[dvips]{epsfig}

\begin{document}

\title[Quantum correlations of Two-Qubit coupled to bath spin.]{ Quantum correlations of Two-Qubit XXZ Heisenberg Chain with
Dzyaloshinsky-Moriya interaction coupled to bath spin as non-Markovian environment}

\author{M. Mahdian, M. Bagherpour Jeddi}
\address{Department of Theoretical Physics and Astrophysics, University of Tabriz, Tabriz 51664, Iran. }
\ead{Mahdian@tabrizu.ac.ir }

\begin{abstract}

We consider the quantum correlations (entanglement and quantum discord) dynamics of two coupled spin qubits with Dzyaloshinsky-Moriya interaction influenced by a locally external magnetic field along $z$-direction and coupled to bath spin-$\frac{1}{2}$ particles as independent non-Markovian environment. We find that with increasing $D_{z}$ and decreasing $J_{z}$, the value of entanglement and quantum discord increase for both antiferromagnetic and ferromagnetic materials. Not that, this growth is more  for the ferromagnetic materials. In addition, we perceive that entanglement and quantum discord decrease with increased temperature and increased coupling constants between reduced system and bath. But, strong quantum correlations within the spins of bath reduce decoherence effects. We discuss about type of the constituent material of the central spins that it can speedup the quantum information processing and as a result, we perceive that one can improve and control the quantum information processing with correct selection of the properties of the reduced system ($J$, $J_{z}$, $D_{z}$).
\end{abstract}
\newpage

\section{Introduction}
In the actual dynamics of any real open quantum system we have faced with the various type of interactions between system and its surrounding environment. These interactions which can lead to decoherence, will cause the transition system from pure quantum states to mixed ones and change the quantum properties, especially quantum correlations. Entanglement and quantum discord which are two different faces of the quantum correlations without classical counterpart, due to their notable features in developing the idea of quantum computers and other quantum information devices have attracted much attention in recent years.
Since both of them have been realized to performing quantum information tasks, the investigation of decoherence dynamics of them is an important emerging field  \cite{4,5,6,7,8,31}. Now, much attention has been paid to the quantum correlation in spin system, such as the Ising model \cite{32} and all kinds of Heisenberg XX, XXZ, XYZ models \cite{33,34,35,36}. Both Ising and XXZ models can be supplemented with a magnetic term, the so called Dzyaloshinskii-Moriya
(DM) interaction. Such antisymmetric exchange interaction arises from the spin-orbit coupling and has many important consequences. Recently, CHEN Yi-Xin and YIN Zhi \cite{40} have performed an interesting investigation on thermal quantum correlations in an anisotropic XXZ model with DM interaction and shown that quantum discord is more robust than entanglement concurrence versus temperature T. They find that the characteristic of quantum discord is unusual in
this system and this possibly offers a potential solution to enhance entanglement of a system.
\\ \\ Just as other quantum systems, spin systems are inevitably influenced by their environment, especially the spin environment. The coupling of spin systems with a spin bath often leads to strong non-Markovian behavior \cite{} which has many physical importance in solid state quantum information processors, such as systems based on the nuclear spin of donors in semiconductors \cite{1,2} or on the electron spin in quantum dots \cite{3}.
However the master equations describing the non-Markovian dynamics are rarely exactly solvable, but recently in Ref. \cite{31} the authors present an exact calculation to obtain the reduced density matrix of two coupled spins in a quantum Heisenberg XY spin environment in the thermodynamic limit at a finite temperature.
For spin systems, much attention has been devoted to the quantum correlations arising in spin chains at thermal equilibrium with their environment.
In this sense, we consider the thermal quantum discord and entanglement within two spin qubit anisotropic XXZ Heisenberg model with antisymmetric DM interaction in the presence the bath spin-$\frac{1}{2}$ particles as environment. Since the dynamics of the system qubit in the model we study is highly non-Markovian and hence it is not expect the traditional Markovian master equations commonly used, we follow the novel technique in Ref. \cite{31}.
In section $2$ we introduce the model Hamiltonian describing two-qubit system with DM interaction coupled to an XY spin chain and derive an analytic formula for the exact solution of non-Markovian dynamics. In section $3$ we review the concept of entanglement and quantum discord and in section $4$ we discuss the effects of the $D_{z}$ (the z-component of the DM interaction), $J_{z}$ (degree of anisotropy),  $J$ ($J>0$ corresponds to the antiferromagnetic case and  $J<0$ corresponds to the ferromagnetic case), $T$ (temperate), $g$ (the bath-system coupling constant) and $g_{0}$ (the inner-bath-spin coupling constant) on entanglement and quantum discord, separately. Our results show that stronger DM interaction and week degree of anisotropy can decreases the disruptive effects of the environment. In addition, increasing the both temperature and the bath-system coupling constant will reduce the amount
of the quantum correlation. Whereas the growth of the inner-bath-spin coupling constant or strong quantum correlations within the spins of bath reduce decoherence and improve the quantum correlations. By comparing the quantum discord with entanglement between two qubits in the model, we have seen that quantum discord is more robust than entanglement versus increasing temperature T and decreasing $D_{z}$. Conclusions are then presented in Section $5$.

%%%%%%%%%%%%%%%%%%%%%%%%%%%%%%%%%%%%%%%%%%%%%%%%%%%%%%%%%%%%%%%%%%%%%%%%%%%%%%%%%%%%%%%%%%%%%%%%%%%%%%%%%%%%%%%%%%%%%%%%%%%%%%%%%%%%%%%%%%%%%%%%%%%%%%%%%%%%%%%%%% %%%%%%%%%%%%%%%%%%%%%%%%%%%%%%%%%%%%%%%%%%%%%%%%%%%%%%%%%%%%%%%%%%%%%%%%%%%%%%%%%%%%%%%%%%%%%%%%%%%%%%%%%%%%%%%%%%%%%%%%%%%%%%%%%%%%%%%%%%%%%%%%%%%%%%%%%%%%%%%%%% %%%%%%%%%%%%%%%%%%%%%%%%%%%%%%%%%%%%%%%%%%%%%%%%%%%%%%%%%%%%%%%%%%%%%%%%%%%%%%%%%%%%%%%%%%%%%%%%%%%%%%%%%%%%%%%%%%%%%%%%%%%%%%%%%%%%%%%%%%%%%%%%%%%%%%%%%%%%%%%%%%
\section{Model and solution  }
The quantum system we consider consists of two spin-$\frac{1}{2}$ anisotropic particles with DM interaction influenced by a locally external magnetic field along $z$-direction coupled to bath spin-$\frac{1}{2}$ particles as environment.
Here the DM interaction is a supplemented magnetic term arising from any interaction of a particle's spin with its motion, which can be represented by the form $\sum_{ij}\vec{D}_{ij}.(\vec{S}_{i}\times \vec{S}_{j})$. Where $\vec{D}$ is the DM vector coupling and the sum is over the pairs of spins. To see the effect of the DM interaction, we choose the $z$-component of the anisotropic parameter $\vec{D}$.
The corresponding Hamiltonian of total system $H_{tot}$ can be written as the sum of Hamiltonians for the system itself $H_{s}$,  the bath spin $H_{b}$ and the coupling between them $H_{sb}$. By considering interaction between the two anisotropic system-spin particles as Heisenberg XXZ model, with DM interaction parameter $D_{z}$, the Hamiltonian of system is given by
\begin{eqnarray}
\hspace{-20mm}
H_{s}
=
\varepsilon (S_{1s}^{z}+S_{2s}^{z})+J_{z}S_{1s}^{z}S_{2s}^{z}+ J (S_{1s}^{+}S_{2s}^{-}+S_{1s}^{-}S_{2s}^{+})+i D_{z}(S_{1s}^{+}S_{2s}^{-}-S_{1s}^{-}S_{2s}^{+}).
\end{eqnarray}
We restrict the interaction between the $N$-components of the bath spin and between the system-bath to such that can be describe as Heisenberg XY model
\begin{eqnarray}
H_{b}&=&\frac{g_{0}}{N}\sum _{j\neq k}^{N}(S_{jb}^{+}S_{kb}^{-}+S_{jb}^{-}S_{kb}^{+}),\cr
H_{sb}&=&\frac{g}{\sqrt{N}}\{(S_{1s}^{+}+S_{2s}^{+})\sum _{j=1}^{N}S_{jb}^{-}+(S_{1s}^{-}+S_{2s}^{-})\sum _{j=1}^{N}S_{jb}^{+}\}.
\end{eqnarray}
For definiteness, the $s$ subscript refers to the system and $b$ the bath spin. Where, the parameter $\varepsilon$ characterize the intensity of the magnetic field applied along the $z$-axis and $D_{z}$ the $z$-component of the DM interaction. The coefficients $J_{z}$, $J$, $g_{0}$ and $g$ correspond to the real coupling constants with $J_{z},J > 0$ for the antiferromagnetic case and $J_{z},J < 0$ for the ferromagnetic case.
By using the collective angular momentum operators $J_{\pm}=\sum _{j=1}^{N}S_{jb}^{\pm}$ and the Holstein-Primakoff transformation as $J_{+}=a^{\dag}\,(\sqrt{N-a^{\dag}a})$ and $J_{-}=(\sqrt{N-a^{\dag}a})\,a$ with $[a,a^{\dag}]=1$, the Hamiltonians of the bath spin $H_{b}$ and interaction $H_{sb}$ can be rewritten as
\begin{eqnarray}
H_{b}&=&g_{0}\{a^{\dag}(1-\frac{a^{\dag}a}{N})a+ \sqrt{1-\frac{a^{\dag}a}{N}} \, aa^{\dag} \, \sqrt{1-\frac{a^{\dag}a}{N}}\}-g_{0},\cr
H_{sb}&=&g\{(S_{1s}^{+}+S_{2s}^{+})\sqrt{1-\frac{a^{\dag}a}{N}}\, a + (S_{1s}^{-}+S_{2s}^{-})a^{\dag}\sqrt{1-\frac{a^{\dag}a}{N}}\}.
\end{eqnarray}
By considering the thermodynamic limit (i.e., $N\rightarrow\infty$) at a finite temperature these above equations are reduced to
\begin{eqnarray}
H_{b}&=&2 g_{0}a^{\dag}a ,\cr
H_{sb}&=&g\{(S_{1s}^{+}+S_{2s}^{+})a + (S_{1s}^{-}+S_{2s}^{-})a^{\dag}\}.
\end{eqnarray}
Observe that, the bath spin is reduced into a single-mode bosonic thermal field with non-Markovian effect on the dynamics of the our system.
Since this thermal field will not remain in a thermal equilibrium state as usually assumed for an environment with very large degrees of freedom, therefor, the master equation approach can not be useful. Following the special operator technique introduced in \cite{31}, by tracing over the degrees of freedom of the environment from density matrix
\begin{eqnarray}
\rho_{tot}(t)=e^{-i H_{tot}\,t}\rho_{tot}(0)e^{i H_{tot}\,t},
\end{eqnarray}
we can catch the exact non-Markovian dynamics of reduced density matrix for the system at arbitrarily finite temperatures.
We assume that the initial density matrix for the total system can be described by a pure and separable state as $\rho_{tot}(0)=\rho_{s}(0)\otimes \rho_{b}$.
Where $\rho_{b}$ refers to the initial density operator of the single-mode bosonic thermal field which at thermal equilibrium is represented by the Boltzmann distribution as $\rho_{b}=\frac{1}{Z}exp[-H_{b}/k_{B}T]$ with the partition function $Z=(1-exp[-2g_{0}/ k_{B}T])^{-1}$, where $k_{B}$ is Boltzmann's constant
which we henceforth set equal to one.
When two spin particles of the system is initially prepared in a normalized state with maximally quantum correlation as
\begin{eqnarray}
|\psi_{s}(0)\rangle= \alpha |00\rangle+ \beta |11\rangle,
\end{eqnarray}
it is easy to check that, the reduced density matrix for the system in the standard basis $|00\rangle, \, |01\rangle, \, |10\rangle, \, |11\rangle$ has the form as
\begin{eqnarray}\label{7}
\hspace{-23mm}
\rho_{s}(t)&=&\frac{1}{Z}\Tr_{b}\,[\,|\alpha|^{2}(e^{-i H_{tot}\,t}|00\rangle e^{-H_{b}/T}\langle 00|e^{i H_{tot}\,t})+ |\beta|^{2}(e^{-i H_{tot}\,t}|11\rangle e^{-H_{b}/T}\langle 11|e^{i H_{tot}\,t}) \cr
\hspace{-23mm}
&+&\alpha\beta^{\ast}(e^{-i H_{tot}\,t}|00\rangle e^{-H_{b}/T}\langle 11|e^{i H_{tot}\,t}) + \alpha^{\ast}\beta(e^{-i H_{tot}\,t}|11\rangle e^{-H_{b}/T}\langle 00|e^{i H_{tot}\,t})\,].
\end{eqnarray}
Calculation of the above equation could be more difficult due to mutual and internal coupled of the system and environment. One way of circumventing this problem introduced in \cite{31} by converting the time evolution equation of the system under the action of the total Hamiltonian into a set of coupled noncommuting operator variable equations. Then by turning the coupled noncommuting operator variable equations into commuting ones via introducing a new set of transformations on the operator variables, the trace over the environmental degrees of freedom can be performed and the exact reduced
dynamics of the system can be obtained.
\\For the case studied here, we can see that
\begin{eqnarray}\label{8}
e^{-i H_{tot}t}|00\rangle= A(t)|00\rangle+B(t)|01\rangle+C(t)|10\rangle+D(t)|11\rangle.
\end{eqnarray}
Note that, the coefficients $A(t),\, B(t), \, C(t)$ and $D(t)$ are functions of operators $a$ and $a^{\dag}$ and do not commute with each other. The Schr\"{o}dinger equation for $e^{-i H_{tot}\,t}|00\rangle $ is
\begin{eqnarray}
\frac{d}{dt}(e^{-i H_{tot}\,t}|00\rangle)=-i H_{tot}(e^{-i H_{tot}t}|00\rangle),
\end{eqnarray}
which by replacing the Eq. (8), transform to 4 coupled first-order differential equations of noncommuting operator variables as
\begin{eqnarray}\label{10}
\hspace{-15mm}
\frac{d}{dt}A(t)&=&-i\{(J_{z}-\varepsilon+2g_{0}a^{\dag}a) A(t)+g a^{\dag}(B(t)+C(t))\},\cr\cr
\hspace{-15mm}
\frac{d}{dt}B(t)&=&-i\{(-J_{z}+2g_{0}a^{\dag}a) B(t)+(J-2i D_{z})C(t)+ g a A(t)+g a^{\dag} D(t) \},\cr\cr
\hspace{-15mm}
\frac{d}{dt}C(t)&=&-i\{(-J_{z}+2g_{0}a^{\dag}a) C(t)+(J+2i D_{z})B(t) +g a A(t)+g a^{\dag} D(t) \},\cr\cr
\hspace{-15mm}
\frac{d}{dt}D(t)&=&-i\{(J_{z}+\varepsilon+2g_{0}a^{\dag}a) D(t)+g a (B(t)+ C(t))\}.
\end{eqnarray}
From Eq. (8), the initial conditions are given by
\begin{eqnarray}
A(0)=1, \,\,\,\,\,\,\, B(0)=C(0)=D(0)=0.
\end{eqnarray}
Since, operator variables which appear in above coupled differential equations do not commute with each other, we cannot use the conventional methods of solving coupled differential equations.
Fortunately, $H_{tot}$ is of an effective Jaynes-Cumming type and it can be block-diagonalized. Therefor, by finding the proper transformations of noncommuting operator variables, we can rewrite the Eq. (10) as the coupled differential equations of complex-number variables.
In this model, we can reach our goal by using transformations as
\begin{eqnarray}
A(t)&=&e^{-2ig_{0}(\hat{n}-1)t}\, A_{1}(t),\cr
B(t)&=&a\,e^{-2ig_{0}(\hat{n}-1)t}\, B_{1}(t),\cr
C(t)&=&a\,e^{-2ig_{0}(\hat{n}-1)t}\, C_{1}(t),\cr
D(t)&=&a a\,e^{-2ig_{0}(\hat{n}-1)t}\, D_{1}(t).
\end{eqnarray}
Observe that, operator variables $A_{1}(t), \, B_{1}(t),\, C_{1}(t) $ and $D_{1}(t)$ are functions of $\hat{n}=a^{\dag}a$ and commute with each other.
Under these transformations the coupled differential equations of noncommuting operator variables change to the coupled differential equations of complex-number and commuting operator variables as
\begin{eqnarray}\label{13}
\hspace{-15mm}
\frac{d}{dt}A_{1}(t)&=&-i\{(J_{z}-\varepsilon+2g_{0}) A_{1}(t)+g\hat{n}(B_{1}(t)+C_{1}(t))\},\cr\cr
\hspace{-15mm}
\frac{d}{dt}B_{1}(t)&=&-i\{-J_{z} B_{1}(t)+(J+2i D_{z})C_{1}(t)+ g  A_{1}(t)+g (\hat{n}-1) D_{1}(t) \},\cr\cr
\hspace{-15mm}
\frac{d}{dt}C_{1}(t)&=&-i\{-J_{z} C_{1}(t)+(J-2i D_{z})B_{1}(t) +g  A_{1}(t)+g (\hat{n}-1) D_{1}(t) \},\cr\cr
\hspace{-15mm}
\frac{d}{dt}D_{1}(t)&=&-i\{(-J_{z}+\varepsilon-2g_{0}) D_{1}(t)+g (B_{1}(t)+ C_{1}(t))\}.
\end{eqnarray}
Notice that, if the total Hamiltonian cannot be block-diagonalized, for example, for the spin anisotropic particles with $x$- or $y$-component DM interaction,
the operator method used here will then not apply to solve the problem exactly.
By solving Eq. (13) in the usual way with initial condition $A_{1}(0)=1$ and $B_{1}(0)=C_{1}(0)=D_{1}(0)=0$, we can evaluate the time evolution for the initial two spins state of $|00\rangle$.
\\A similar analysis as above can be made if the two spins is initially prepared in $|11\rangle$ state. Let
\begin{eqnarray}
e^{-i H_{tot}t}|11\rangle= \tilde{A}(t)|00\rangle+\tilde{B}(t)|01\rangle+\tilde{C}(t)|10\rangle+\tilde{D}(t)|11\rangle.
\end{eqnarray}
For this case, the proper transformations of noncommuting operator variables have the form as
\begin{eqnarray}
\tilde{A}(t)&=&a^{\dag}a^{\dag}\,e^{-2ig_{0}(\hat{n}+1)t}\, \tilde{A}_{1}(t),\cr
\tilde{B}(t)&=&a^{\dag}\,e^{-2ig_{0}(\hat{n}+1)t}\, \tilde{B}_{1}(t),\cr
\tilde{C}(t)&=&a^{\dag}\,e^{-2ig_{0}(\hat{n}+1)t}\, \tilde{C}_{1}(t),\cr
\tilde{D}(t)&=&e^{-2ig_{0}(\hat{n}+1)t}\, \tilde{D}_{1}(t).
\end{eqnarray}
Insertion of these transformations in to Eq. (10) yield
\begin{eqnarray}\label{16}
\hspace{-15mm}
\frac{d}{dt}\tilde{A}_{1}(t)&=&-i\{(J_{z}-\varepsilon+2g_{0}) \tilde{A}_{1}(t)+g\hat{n}(\tilde{B}_{1}(t)+\tilde{C}_{1}(t))\},\cr\cr
\hspace{-15mm}
\frac{d}{dt}\tilde{B}_{1}(t)&=&-i\{-J_{z} \tilde{B}_{1}(t)+(J+2i D_{z})\tilde{C}_{1}(t)+ g(\hat{n}+2)  \tilde{A}_{1}(t)+g \tilde{D}_{1}(t) \},\cr\cr
\hspace{-15mm}
\frac{d}{dt}\tilde{C}_{1}(t)&=&-i\{-J_{z} \tilde{C}_{1}(t)+(J-2i D_{z})\tilde{B}_{1}(t) +g(\hat{n}+2)  \tilde{A}_{1}(t)+g \tilde{D}_{1}(t) \},\cr\cr
\hspace{-15mm}
\frac{d}{dt}\tilde{D}_{1}(t)&=&-i\{(J_{z}+\varepsilon-2g_{0}) \tilde{D}_{1}(t)+g(\hat{n}+1) (\tilde{B}_{1}(t)+ \tilde{C}_{1}(t))\}.
\end{eqnarray}
By solving these coupled differential equations of complex-number and commuting operator variables in the usual way with initial condition $\tilde{A}_{1}(0)=\tilde{B}_{1}(0)=\tilde{C}_{1}(0)=0$ and $\tilde{D}_{1}(0)=1$, we can evaluate the time evolution for $|11\rangle$.
From the results of Eqs. (13) and (16), the reduced density matrix Eq. (7) in the representation spanned by the two-qubit
product states $|00\rangle=|0\rangle_{1}\otimes|0\rangle_{2}$, $|01\rangle=|0\rangle_{1}\otimes|1\rangle_{2}$, $|10\rangle=|1\rangle_{1}\otimes|0\rangle_{2}$ and $|11\rangle=|1\rangle_{1}\otimes|1\rangle_{2}$ can be written as
\begin{eqnarray}\label{17}
\rho_{s}(t)=\left(\matrix{\rho_{11}&0&0&\rho_{14}\cr 0&\rho_{22}&\rho_{23}&0 \cr 0&\rho_{32}&\rho_{33}&0 \cr \rho_{14}^{*}&0&0&\rho_{44} }\right),
\end{eqnarray}
with
\begin{eqnarray}
\hspace{-15mm}
\rho_{11}&=&\frac{1}{Z}\sum_{n=0}^{\infty} (\,|\alpha|^{2}\,\tilde{A}_{1}(t)\,\tilde{A}_{1}^{\dag}(t)+(n+1)(n+2)|\beta|^{2}\,A_{1}(t)\,A_{1}^{\dag}(t)\,)\,e^{-2ng_{0}/T}\cr
\hspace{-15mm}
\rho_{14}&=&\frac{e^{4igt}}{Z}\sum_{n=0}^{\infty}\alpha\beta^{*}\, \tilde{A}_{1}(t)\, D_{1}^{\dag}\,e^{-2ng_{0}/T}\cr
\hspace{-15mm}
\rho_{22}&=&\rho_{23}=\rho_{32}=\rho_{33}
=\frac{1}{Z}\{|\beta|^{2}\,B_{1}(t)\,B_{1}^{\dag}(t)\cr
\hspace{-15mm}
&+&\sum_{n=1}^{\infty} (\,|\alpha|^{2}\,n\,\tilde{B}_{1}(t)\,\tilde{B}_{1}^{\dag}(t)+(n+1)|\beta|^{2}\,B_{1}(t)\,B_{1}^{\dag}(t)\,)\,e^{-2ng_{0}/T}\}\cr
\hspace{-15mm}
\rho_{44}&=&\frac{1}{Z} \{|\beta|^{2}\,D_{1}(t)\,D_{1}^{\dag}(t)(1+e^{-2g_{0}/T})\cr
\hspace{-15mm}
&+&\sum_{n=2}^{\infty} (\,|\alpha|^{2}\,n(n-1)\,\tilde{D}_{1}(t)\,\tilde{D}_{1}^{\dag}(t)+|\beta|^{2}\,D_{1}(t)\,D_{1}^{\dag}(t)\,)\,e^{-2ng_{0}/T}\}
\end{eqnarray}
Unfortunately, achieved analytic expression of solutions of Eqs. (13) and (16) has no compact form and, thus, we do not show it here explicitly.

%%%%%%%%%%%%%%%%%%%%%%%%%%%%%%%%%%%%%%%%%%%%%%%%%%%%%%%%%%%%%%%%%%%%%%%%%%%%%%%%%%%%%%%%%%%%%%%%%%%%%%%%%%%%%%%%%%%%%%%%%%%%%%%%%%%%%%%%%%%%%%%%%%%%%%%%%%%%%%%%%% %%%%%%%%%%%%%%%%%%%%%%%%%%%%%%%%%%%%%%%%%%%%%%%%%%%%%%%%%%%%%%%%%%%%%%%%%%%%%%%%%%%%%%%%%%%%%%%%%%%%%%%%%%%%%%%%%%%%%%%%%%%%%%%%%%%%%%%%%%%%%%%%%%%%%%%%%%%%%%%%%% %%%%%%%%%%%%%%%%%%%%%%%%%%%%%%%%%%%%%%%%%%%%%%%%%%%%%%%%%%%%%%%%%%%%%%%%%%%%%%%%%%%%%%%%%%%%%%%%%%%%%%%%%%%%%%%%%%%%%%%%%%%%%%%%%%%%%%%%%%%%%%%%%%%%%%%%%%%%%%%%%%

\section{Quantum correlations}
The role of entanglement in developing the idea of quantum computers \cite{23} and sending information in novel ways, such as quantum teleportation or quantum cryptography, has turned this intrinsic principle of quantum mechanics into one of the most prolific topics. To investigate the entanglement dynamics of the our bipartite system, we apply Wootters concurrence \cite{24}. The concurrence can be calculated explicitly from the time dependent density matrix $\rho_{s}(t)$
of the two spins
\begin{eqnarray}
C(\rho_{s}(t))=\max\{0, \lambda_{1}-\lambda_{2}-\lambda_{3}-\lambda_{4}\},
\end{eqnarray}
where the quantities $\lambda_{i}$ are the square roots of the eigenvalues of the matrix $\vartheta=\rho_{s}(t)(\sigma_{y}\otimes \sigma_{y})\rho_{s}^{*}(t)(\sigma_{y}\otimes \sigma_{y})$, arranged in decreasing order. Here $\rho_{s}^{*}(t)$ means the complex conjugation of $\rho_{s}(t)$, and $\sigma_{y}$ is the Pauli matrix.
The concurrence varies from zero for a separable state to one for a maximally entangled state.
The quantum state Eq. (17) are entangled if and only if either $\rho_{22}\rho_{33} < |\rho_{14}|^{2}$ or $\rho_{11}\rho_{44}< |\rho_{23}|^{2}$.
Both conditions cannot hold simultaneously \cite{10}.
The entanglement of this state is obtained as
\begin{eqnarray}
C(\rho_{s}(t))= \sqrt{\rho_{11}\rho_{44}}+|\rho_{14}|-|\sqrt{\rho_{11}\rho_{44}}-|\rho_{14}||-2\rho_{22}.
\end{eqnarray}
However, concurrence is not the only type of quantum correlations.
The different kind of quantum correlations than entanglement, so called quantum discord, has interesting and significant applications in quantum information processing.
Quantum discord is not always larger than entanglement
\cite{41,42}.
This indicates that discord is not simply a sum of entanglement and some other nonclassical correlation.
The relation between quantum discord, entanglement, and classical correlation even for the simplest case of two entangled qubits is not yet clear.
In a bipartite quantum state with density matrix operator $\rho_{s}(t)$, that is inclusive of two part and has the form as two-qubit X states \cite{43}, quantum discord is introduced as the difference between the total correlation and the classical correlation  with the following expression
\begin{eqnarray}
D(\rho_{s}(t))= I(\rho_{s}(t))- C(\rho_{s}(t)).
\end{eqnarray}
There
\begin{eqnarray}
I(\rho_{s}(t))= S(\rho_{1s})+S(\rho_{2s})- \sum_{j=1}^{4}\gamma_{j}\log_{2}\gamma_{j},
\end{eqnarray}
is the quantum mutual information. In this notation $\gamma_{j}$ are the eigenvalues of the density matrix $\rho_{s}(t)$
\begin{eqnarray}
\gamma_{1}&=& \frac{1}{2}(\rho_{11}+\rho_{44}+ \sqrt{(\rho_{11}-\rho_{44})^{2}+4|\rho_{14}|^{2}}),\cr
\gamma_{2}&=& \frac{1}{2}(\rho_{11}+\rho_{44}- \sqrt{(\rho_{11}-\rho_{44})^{2}+4|\rho_{14}|^{2}}),\cr
\gamma_{3}&=& \frac{1}{2}(\rho_{22}+\rho_{33}+ \sqrt{(\rho_{22}-\rho_{33})^{2}+4|\rho_{23}|^{2}}),\cr
\gamma_{4}&=& \frac{1}{2}(\rho_{22}+\rho_{33}- \sqrt{(\rho_{22}-\rho_{33})^{2}+4|\rho_{23}|^{2}}),
\end{eqnarray}
and
\begin{eqnarray}
\hspace{-15mm}
S(\rho_{1s})&=& -[(\rho_{11}+\rho_{22})\log_{2}(\rho_{11}+\rho_{22})+(\rho_{33}+\rho_{44})\log_{2}(\rho_{22}+\rho_{33})],\cr
\hspace{-15mm}
S(\rho_{2s})&=& -[(\rho_{11}+\rho_{33})\log_{2}(\rho_{11}+\rho_{33})+(\rho_{22}+\rho_{44})\log_{2}(\rho_{22}+\rho_{44})],
\end{eqnarray}
are the von Neumann entropy of $\rho_{1s}$ and $\rho_{2s}$, the marginal states of $\rho_{s}(t)$.
In order to computing the classical correlation $C(\rho_{s}(t))$, we must perform a suitable projection measure on subsystem $2$. After this, the state
$\rho_{s}(t)$ will change to the ensemble $\{\rho_{i}; p_{i}\}$ ($i=0,1$) with
\begin{eqnarray}
\rho_{i}&=&\frac{1}{p_{i}}(I\otimes V |i\rangle\langle i| V^{\dagger})\rho_{s}(t)(I\otimes V |i\rangle\langle i| V^{\dagger}),\cr
p_{i}&=&\Tr[(I\otimes V |i\rangle\langle i| V^{\dagger})\rho_{s}(t)(I\otimes V |i\rangle\langle i| V^{\dagger})].
\end{eqnarray}
Where $V\in SU(2)$ is a unitary operator with unit determinant. The ensemble $\{\rho_{i}; p_{i}\}$ can be characterized by
their eigenvalues as
\begin{eqnarray}
\alpha_{\pm}(\rho_{0})&=& \frac{1}{2}\,(1\pm \, \frac{1}{p_{0}}\sqrt{[(\rho_{11}-\rho_{33})k+(\rho_{22}-\rho_{44})l]^{2}+\Theta}\,\,\,),\cr
\beta_{\pm}(\rho_{1})&=& \frac{1}{2}\,(1\pm \, \frac{1}{p_{1}}\sqrt{[(\rho_{11}-\rho_{33})l+(\rho_{22}-\rho_{44})k]^{2}+\Theta}\,\,\,),
\end{eqnarray}
with
\begin{eqnarray}
\hspace{-15mm}
\Theta=4kl(|\rho_{14}|^{2}+|\rho_{23}|^{2}+2Re (\rho_{14}\rho_{23}))-16 m Re(\rho_{14}\rho_{23}))+16 n Im (\rho_{14}\rho_{23})),
\end{eqnarray}
and their corresponding probabilities as
\begin{eqnarray}
p_{0}&=&(\rho_{11}+\rho_{33})k+(\rho_{22}+\rho_{44})l\, ,\cr
p_{1}&=&(\rho_{11}+\rho_{33})l+(\rho_{22}+\rho_{44})k\,.
\end{eqnarray}
The parameters $m$, $n$, $k$ and $l$ are dependent to the projection of the von Neumann measurement $V$ on the Bloch sphere. By rewriting this operator as $V = tI+ i\vec{Y}.\vec{\sigma} $ with $t, y_{1}, y_{2}, y_{3}\in R$ and $t^{2}+y_{1}^{2}+y_{2}^{2}+y_{3}^{2}=1$ we have
\begin{eqnarray}
\hspace{-22mm}
m=(t y_{1}+y_{2}y_{3})^{2},\,\,\,\,\,\,\, n=(t y_{2}-y_{1}y_{3})(t y_{1}+y_{2}y_{3}),\,\,\,\,\,\, k=t^{2}+y_{3}^{2},\,\,\,\,\,\, l=1-k.
\end{eqnarray}
The classical correlation is obtained as $C(\rho_{s}(t))=S(\rho_{1s}) - \min_{\{V\}}(p_{0}\,S(\rho_{0})+p_{1}\,S(\rho_{1}))$.
Therefore, to calculate the classical correlation and consequently quantum discord, we have to minimize the quantum conditional entropy  $(p_{0}\,S(\rho_{0})+p_{1}\,S(\rho_{1}))$ with respect to the von Neumann measurements.
In our model the minimization above expression attained in $k=l=1/2$ and $m=n=0$.

%%%%%%%%%%%%%%%%%%%%%%%%%%%%%%%%%%%%%%%%%%%%%%%%%%%%%%%%%%%%%%%%%%%%%%%%%%%%%%%%%%%%%%%%%%%%%%%%%%%%%%%%%%%%%%%%%%%%%%%%%%%%%%%%%%%%%%%%%%%%%%%%%%%%%%%%%%%%%%%%%
%%%%%%%%%%%%%%%%%%%%%%%%%%%%%%%%%%%%%%%%%%%%%%%%%%%%%%%%%%%%%%%%%%%%%%%%%%%%%%%%%%%%%%%%%%%%%%%%%%%%%%%%%%%%%%%%%%%%%%%%%%%%%%%%%%%%%%%%%%%%%%%%%%%%%%%%%%%%%%%%%
%%%%%%%%%%%%%%%%%%%%%%%%%%%%%%%%%%%%%%%%%%%%%%%%%%%%%%%%%%%%%%%%%%%%%%%%%%%%%%%%%%%%%%%%%%%%%%%%%%%%%%%%%%%%%%%%%%%%%%%%%%%%%%%%%%%%%%%%%%%%%%%%%%%%%%%%%%%%%%%%%

\section{Quantum correlations of two coupled spins in spin bath}
In this section, we will discuss the time variations of the quantum correlations for antiferromagnetic ($J>0$) and ferromagnetic ($J<0$) in our model.
Since the explicit expressions of entanglement and the quantum discord are very complicated, here we skip the details and give our results
in terms of figures.
\\ \\In order to demonstrate the properties of different DM coupling parameter $D_{z}$ on quantum correlations in antiferromagnetic case, in Fig. 1(a) the thermal entanglement and in Fig. 1(b) the thermal quantum discord versus time are plotted for different values of $D_{z}$. For both figures the coupling constants are $J=2.0$  and $J_{z}=1.0$ and other parameters are $\varepsilon=0.5, \, g_{0}=g=1.0$ and $T=2.0$. From these figures, it's obvious that the vanishing rate of the entanglement and the quantum discord decreases as the DM interaction parameter $D_{z}$ increases. This phenomenon indicates that, the decays of the quantum correlations due to interaction with spin bath can be compensated by tuning the DM interaction. In addition, it is easy to see that both entanglement and the quantum discord are recovered in the long time as we expected from non-Markovian dynamics. The comparison between these two figures shows that the quantum discord is more robust than entanglement under destroyed conditions.
\\ \\ Similarly, Fig. 2 shows the characteristics of different $D_{z}$ on quantum correlations in ferromagnetic case ($J=-2.0$). Fig. 2(a) illustrates the thermal entanglement and Fig. 2(b) illustrates the thermal quantum discord versus time with $J_{z}=1.0$, $\varepsilon=0.5, \, g_{0}=g=1.0$ and $T=2.0$.
In this condition, the quantum correlations decrease with gentle slope than antiferromagnetic case. As is indicated in Figs. 2(a) and 2(b)
quantum correlations in ferromagnetic case is more robust than antiferromagnetic case versus increasing the values of the DM interaction.
\\ \\ In Figs. 3 and 4 to exhibit the effects of the coupling constant $J_{z}$ on quantum correlations in antiferromagnetic ($J=2.0$), the thermal entanglement in Figs. 3(a) and 4(a), and the thermal quantum discord in Figs. 3(b) and 4(b) versus time for different values of $D_{z}$ are illustrated when $\varepsilon=0.5, \, g_{0}=g=1.0$ and $T=2.0$. In Fig. 3 the coupling constant $J_{z}$ has a lesser value than the same in Figs. 1 and 4. The comparison between these three figures shows that increasing the  coupling parameter $J_{z}$ can enhance the fluctuations of the quantum correlations.
It seems that increasing the $J_{z}$ will decrease the effects of the DM interaction. In one side, the curves of the higher value of $J_{z}$ are closer than of the lower ones. On the other side, the diminution of the both entanglement and the quantum discord in the long time scale for the higher values of $D_{z}$ is more in system with strong coupling.  So as a result, the diminution of the quantum correlations can be compensated both by increasing the $D_{z}$ and by decreasing $J_{z}$ of the reduced system.
\\ \\ The effects of temperature T on quantum correlations are illustrated in Fig. 5. Where entanglement and the quantum discord versus time and temperature $T$ with $J=2.0$, $J_{z}=1.0$, $\varepsilon=0.5, \, g_{0}=g=1.0$ and $D_{z}=2.0$ are depicted in Figs. 5(a) and 5(b), respectively. We found that in the very small temperature, both entanglement and quantum discord have their maximum value, but the increased temperature  can decrease them. This decreasing of quantum correlations  can be explained as follows. In the very small temperature, no excitation for the spins of the bath will exist and the bath is in a thoroughly polarized state with all spins down. On the other word, the effects of decoherence induced by the bath are not strong enough to weaken the quantum correlations, but in the high temperature, the spin orientation of the  bath becomes the very disorder, consequently effects of thermal fluctuation of the spin bath overcome on the effects of the quantum fluctuations (due to the $D_{z}$ interaction). With intensity decrease of the connection  between bath and the reduced system, the role of mediation of the spin bath weaken, therefore two qubits of the reduced system can't exchange information with together and the value of quantum correlations reduced \cite{27}. Not that, the quantum discord unlike entanglement never be zero. As a result, in the such as open quantum systems that the inner-bath-spin coupling constant is the strong, one can with right tuning $J_{z}$ and $D_{z}$ improve the both entanglement and quantum discord.

\section{Conclusions}
In summary, we have investigated the quantum discord  and entanglement within two-qubit anisotropic XXZ Heisenberg model with antisymmetric DM interaction tunable parameter, such as  $D_{z}$, influenced by a locally external magnetic field along $z$-direction in the presence spin bath. We obtain that with decreasing $J_{z}$ and increasing $D_{z}$, the value of entanglement and the quantum discord increase for the  antiferromagnetic.
In the same conditions for the ferromagnetic the quantum correlations decrease with gentle slope than antiferromagnetic case. In addition, we obtain that increasing the values of T (temperature)  will reduce the amount of the both entanglement and quantum discord. Finally, we propose that in the such as open quantum systems that we can't reduce the devastating effects of T and $g$, we can  with right tuning $J_{z}$ and $D_{z}$ or exchanging the constituent material of the central spins improve the both quantities of the entanglement and the quantum discord.

\newpage
\begin{figure}
\centering
\includegraphics[width=390 pt]{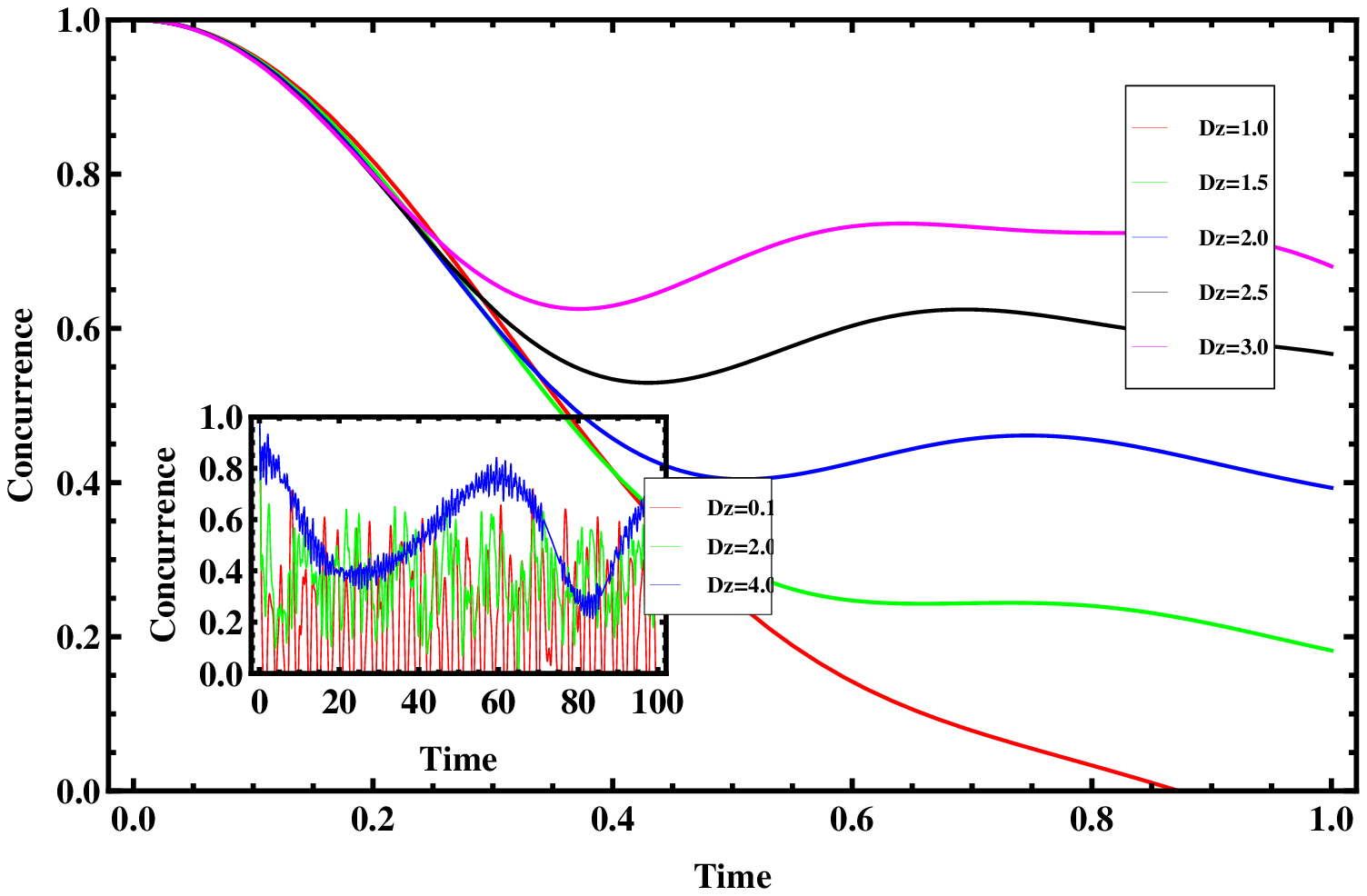}
\caption{}
\label{FIG. 1(a)}
\end{figure}
\textbf{Figure Caption}
\itemize{}
\item Fig. 1(a): (Color online) The time evolution of the entanglement for initial two-qubit state $|\psi_{s}(0)\rangle=\frac{1}{\surd2}( |00\rangle+ |11\rangle)$  for the  antiferromagnetic case $J=2.0$ and $J_{z}=1.0$. Other parameters are $\varepsilon=0.5, \,g_{0}=g =1.0$ and $T=2.0$.

\newpage
\begin{figure}
\centering
\includegraphics[width=390 pt]{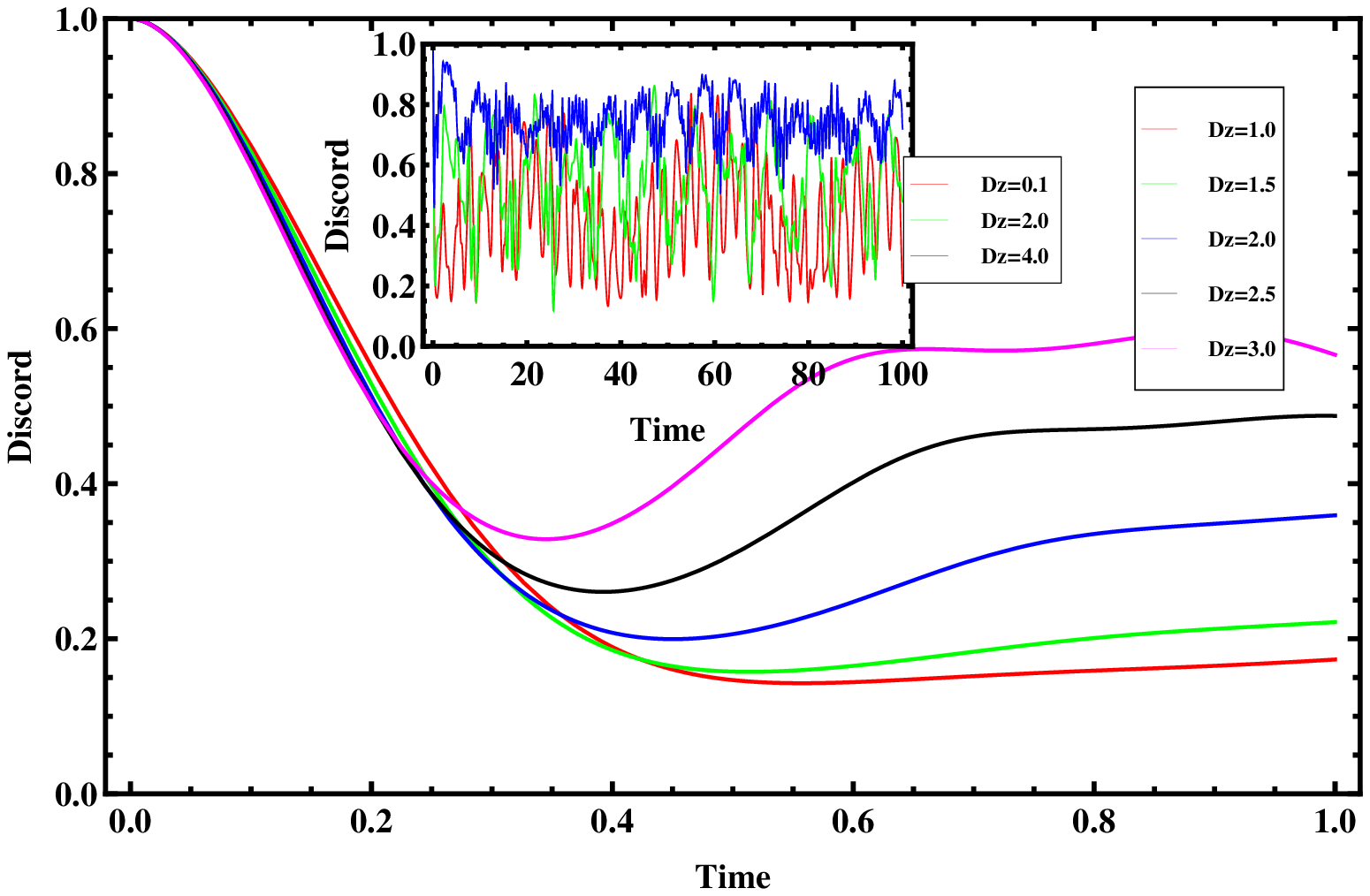}
\caption{}
\label{FIG. 1(b)}
\end{figure}
\textbf{Figure Caption}
\item Fig. 1(b): (Color online) The time evolution of the quantum discord for initial two-qubit state $|\psi_{s}(0)\rangle=\frac{1}{\surd2}( |00\rangle+ |11\rangle)$  for the  antiferromagnetic case $J=2.0$ and $J_{z}=1.0$. Other parameters are $\varepsilon=0.5, \,g_{0}=g =1.0$ and $T=2.0$.

\newpage
\begin{figure}
\centering
\includegraphics[width=445 pt]{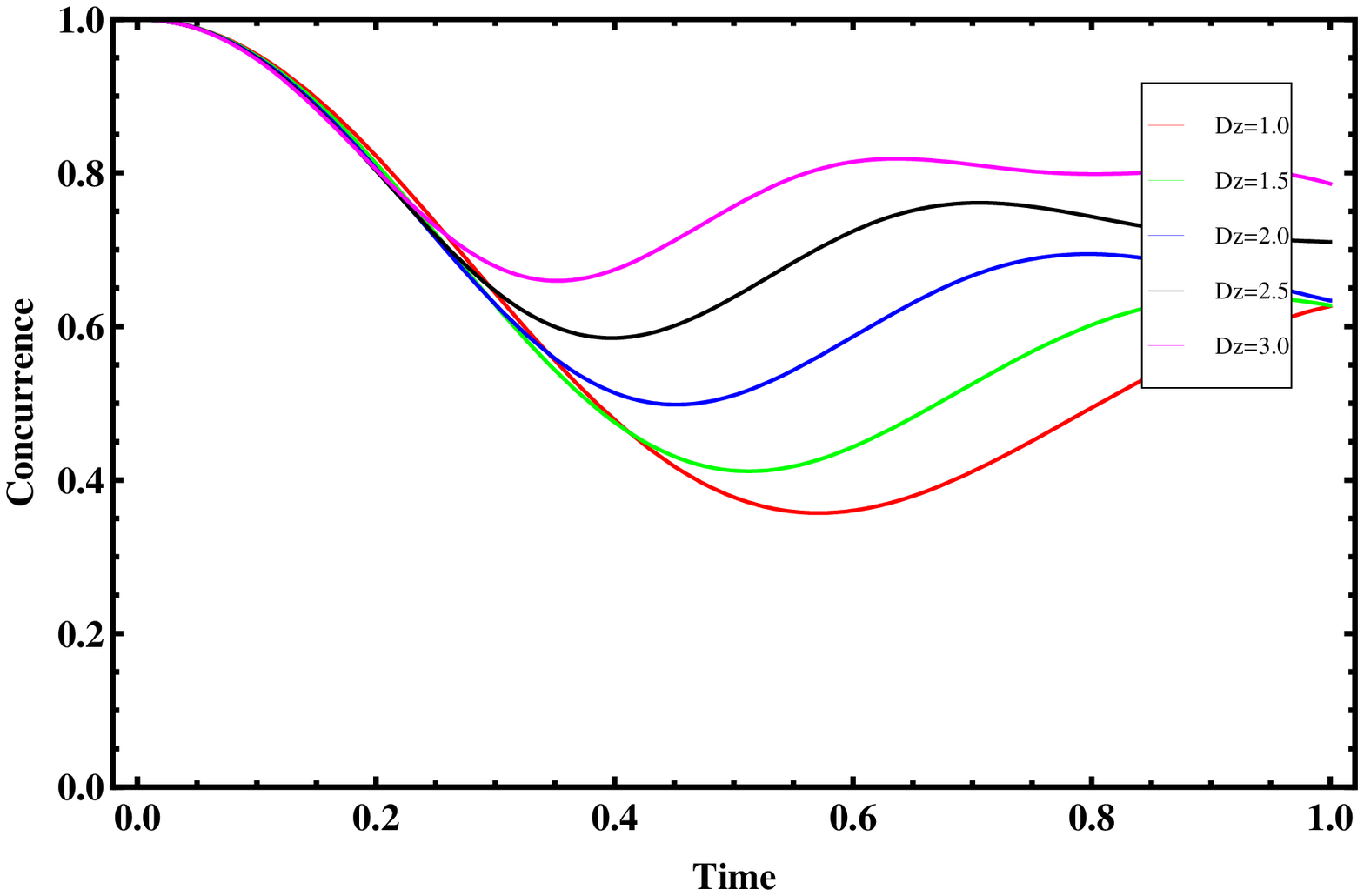}
\caption{}
\label{FIG. 2(a)}
\end{figure}
\textbf{Figure Caption}
\item Fig. 2(a):(Color online) The time evolution of the entanglement for initial two-qubit state $|\psi_{s}(0)\rangle=\frac{1}{\surd2}( |00\rangle+ |11\rangle)$  for the  ferromagnetic case $J=-2.0$ and $J_{z}=0.0$. Other parameters are $\varepsilon=0.5, \,g_{0}=g =1.0$ and $T=2.0$.

\newpage
\begin{figure}
\centering
\includegraphics[width=445 pt]{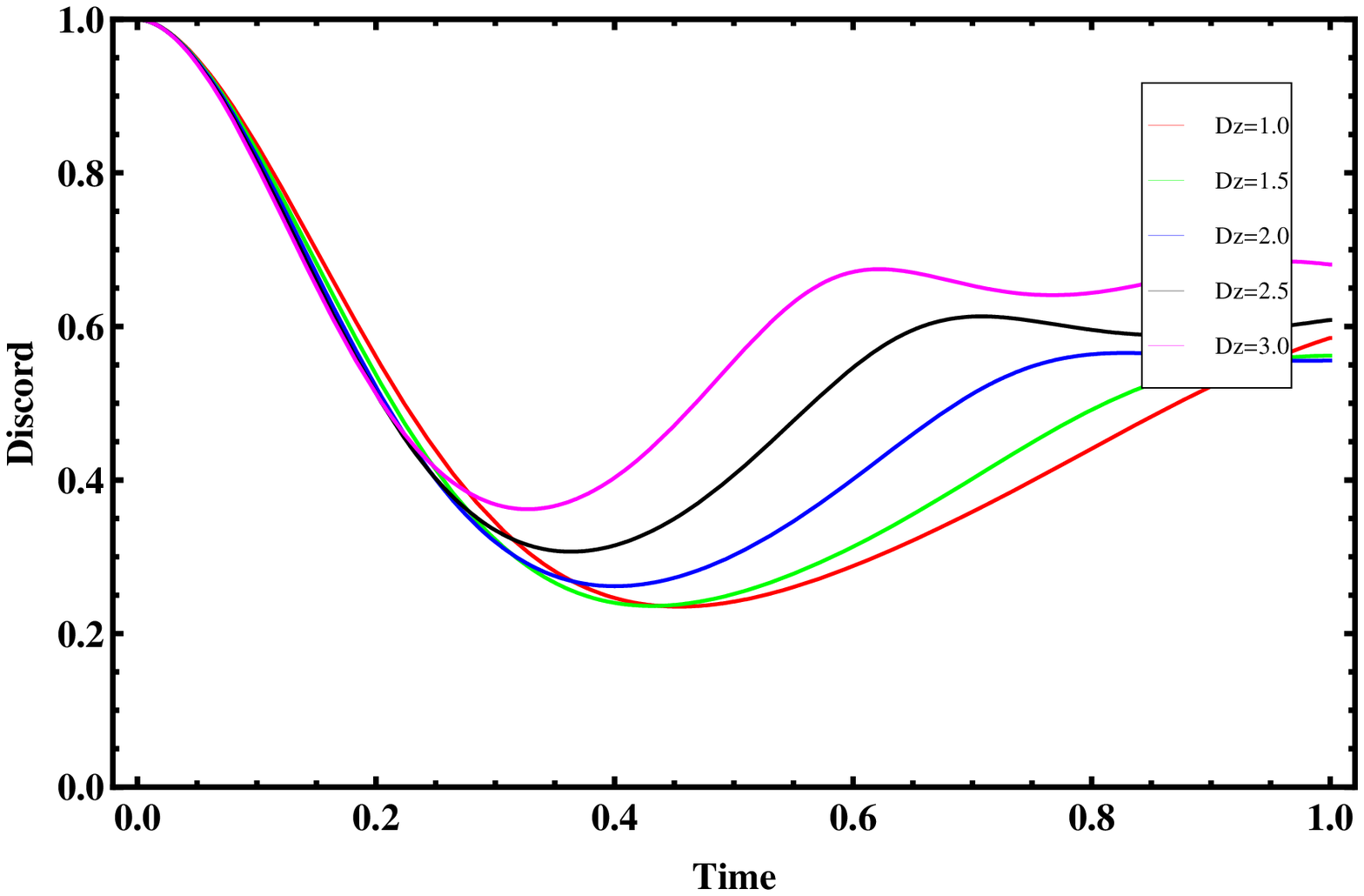}
\caption{}
\label{FIG. 2(b)}
\end{figure}
\textbf{Figure Caption}
\item FIG. 2(b): (Color online) The time evolution of the quantum discord for initial two-qubit state $|\psi_{s}(0)\rangle=\frac{1}{\surd2}( |00\rangle+ |11\rangle)$  for the  ferromagnetic case $J=-2.0$ and $J_{z}=0.0$. Other parameters are $\varepsilon=0.5, \,g_{0}=g =1.0$ and $T=2.0$.

\newpage
\begin{figure}
\centering
\includegraphics[width=445 pt]{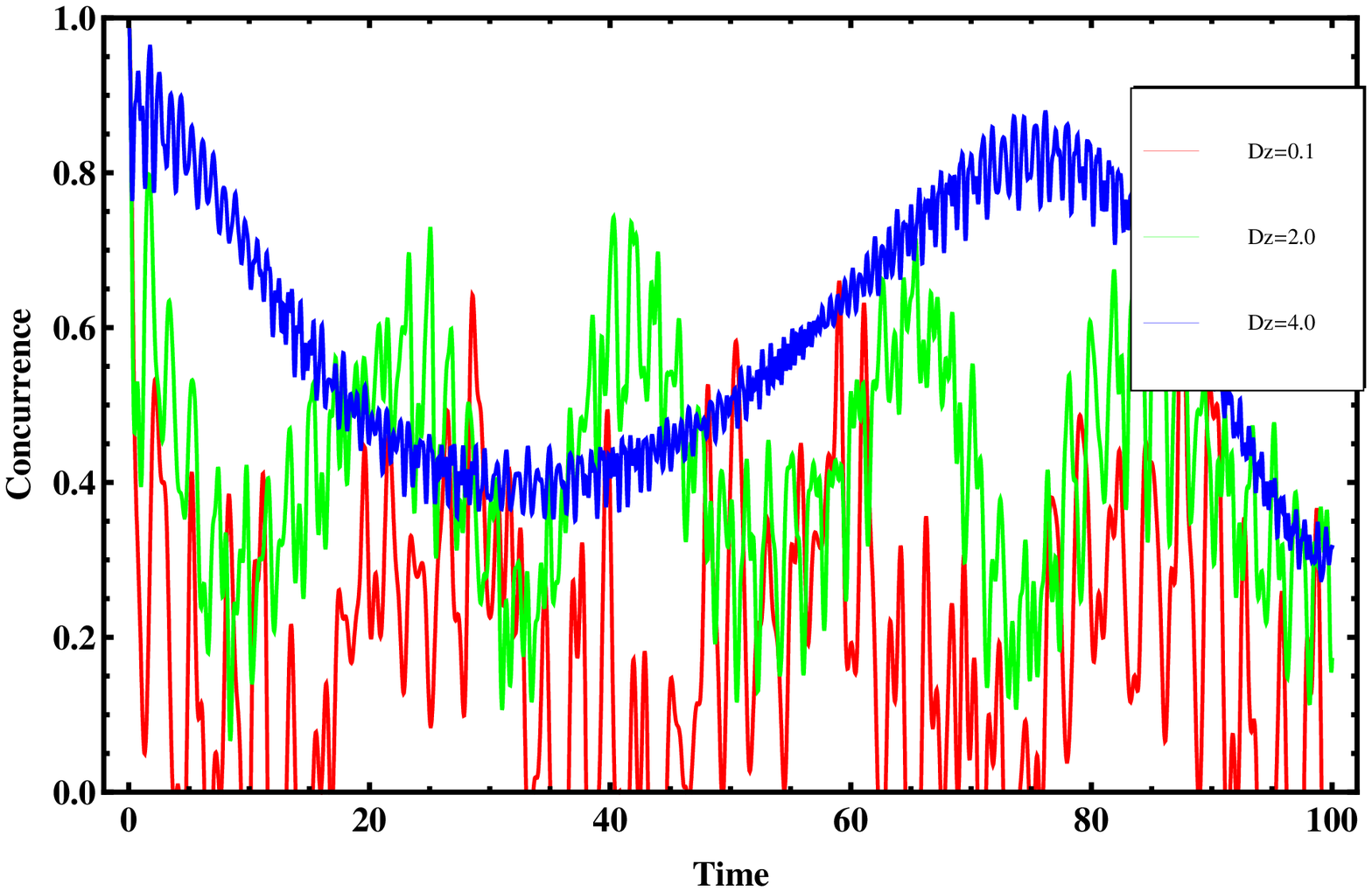}
\caption{}
\label{FIG. 3(a)}
\end{figure}
\textbf{Figure Caption}
\item Fig. 3(a): (Color online) The time evolution of the entanglement for initial two-qubit state $|\psi_{s}(0)\rangle=\frac{1}{\surd2}( |00\rangle+ |11\rangle)$  for the  antiferromagnetic case $J=2.0$ and $J_{z}=2.0$. Other parameters are $\varepsilon=0.5, \,g_{0}=g =1.0$ and $T=2.0$.
\newpage
\begin{figure}
\centering
\includegraphics[width=390 pt]{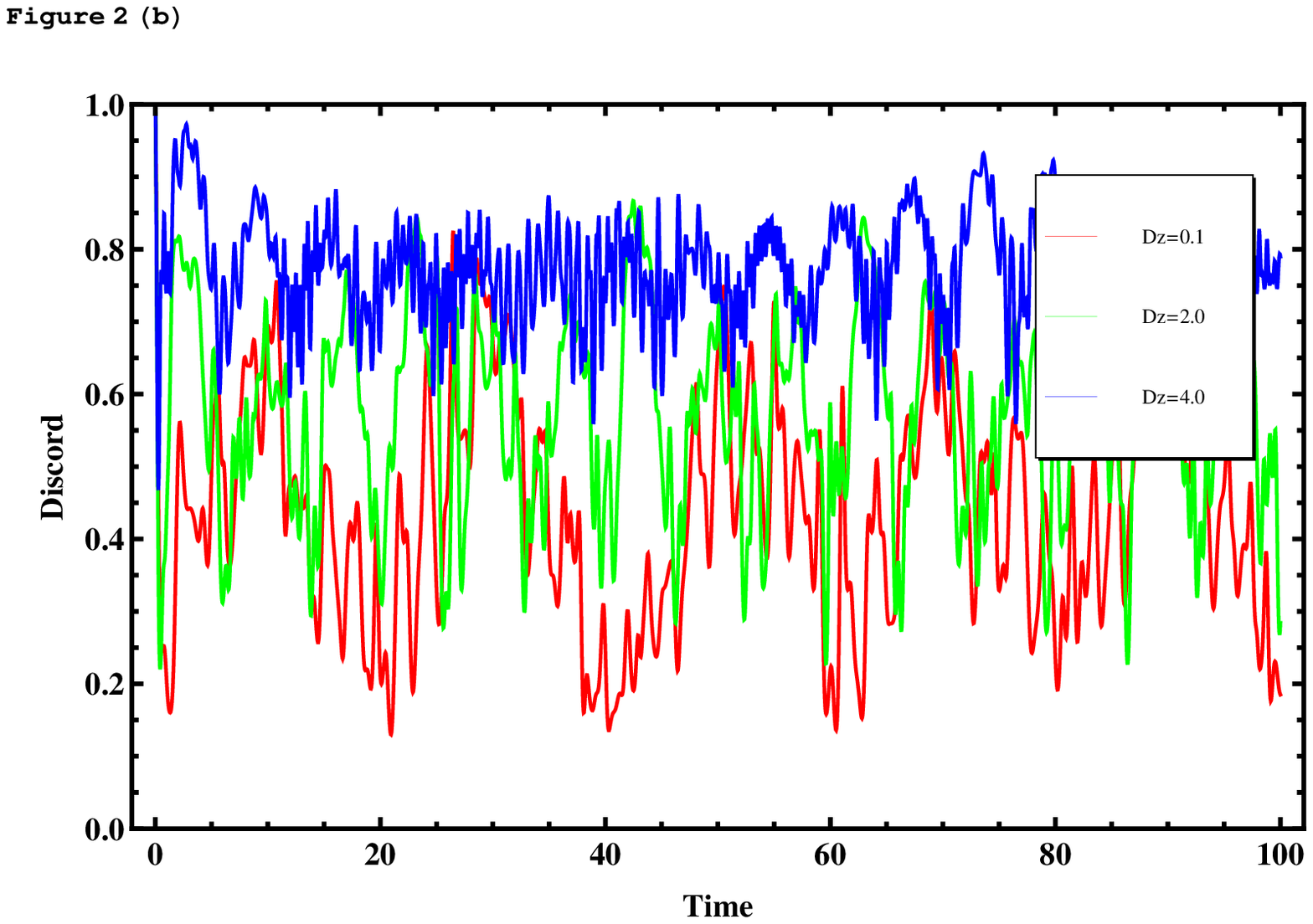}
\caption{}
\label{FIG. 3(b)}
\end{figure}
\textbf{Figure Caption}
\item Fig. 3(b): (Color online) The time evolution of the quantum discord for initial two-qubit state $|\psi_{s}(0)\rangle=\frac{1}{\surd2}( |00\rangle+ |11\rangle)$  for the  antiferromagnetic case $J=2.0$ and $J_{z}=2.0$. Other parameters are $\varepsilon=0.5, \,g_{0}=g =1.0$ and $T=2.0$.

\newpage
\begin{figure}
\centering
\includegraphics[width=445 pt]{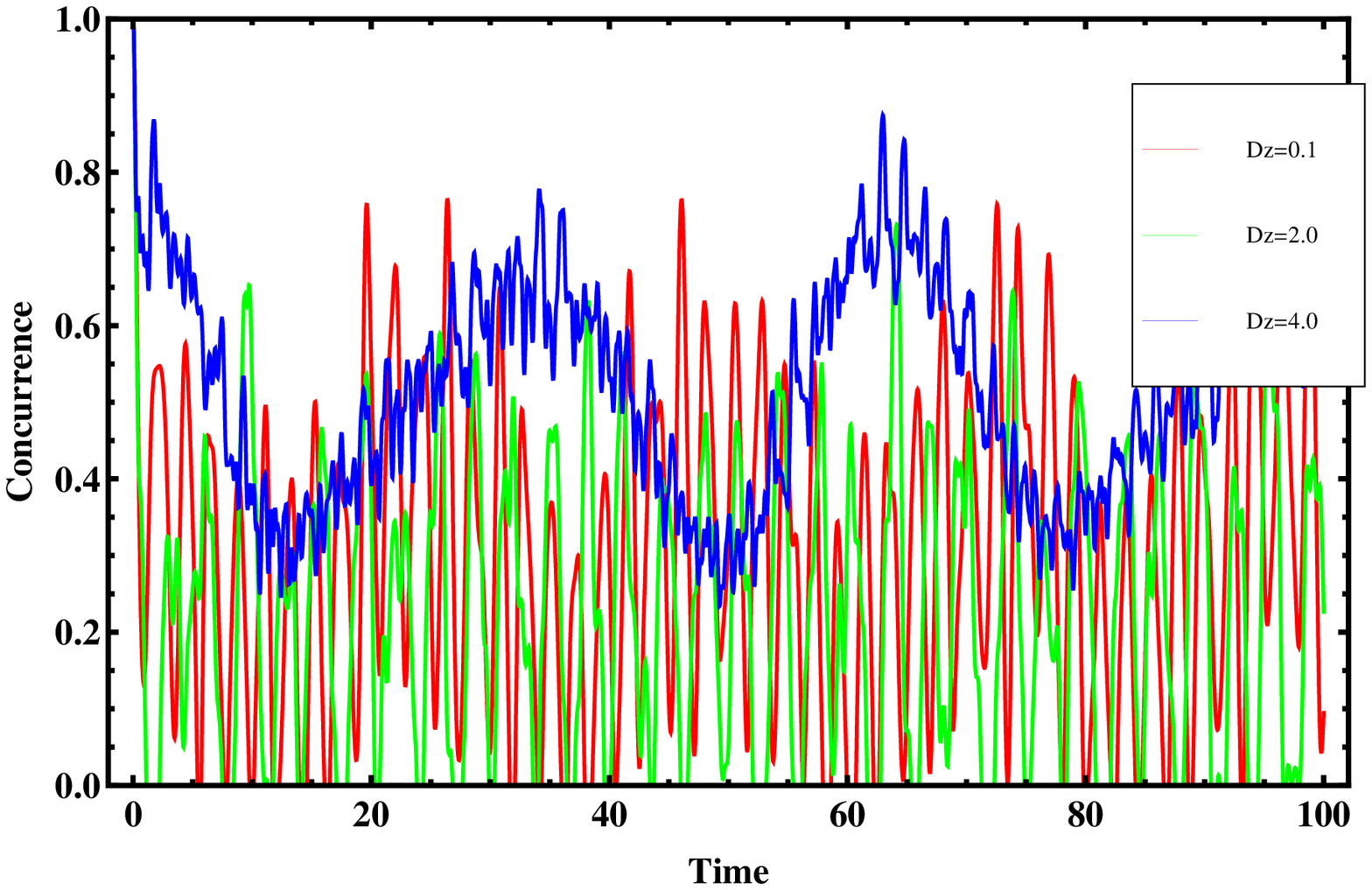}
\caption{}
\label{FIG. 4(a)}
\end{figure}
\textbf{Figure Caption}
\item Fig. 4(a): (Color online) The time evolution of the entanglement for initial two-qubit state $|\psi_{s}(0)\rangle=\frac{1}{\surd2}( |00\rangle+ |11\rangle)$  for the ferromagnetic case $J=2.0$ and $J_{z}=1.0$. Other parameters are $\varepsilon=0.5, \,g_{0}=g =1.0$ and $T=2.0$.

\newpage
\begin{figure}
\centering
\includegraphics[width=445 pt]{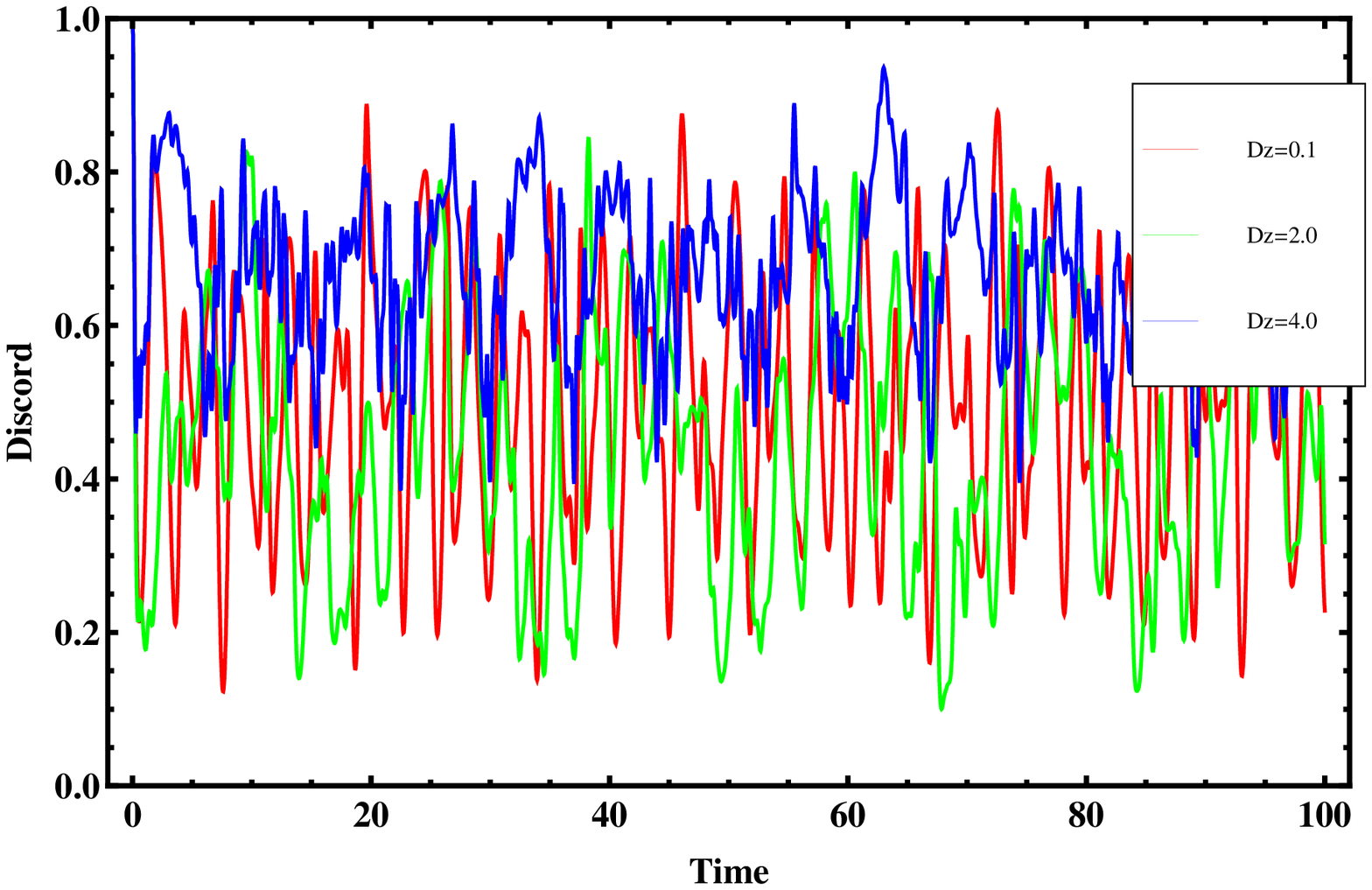}
\caption{}
\label{FIG. 4(b)}
\end{figure}
\textbf{Figure Caption}
\item Fig. 4(b): (Color online) The time evolution of the quantum discord for initial two-qubit state $|\psi_{s}(0)\rangle=\frac{1}{\surd2}( |00\rangle+ |11\rangle)$  for the ferromagnetic case $J=2.0$ and $J_{z}=1.0$. Other parameters are $\varepsilon=0.5, \,g_{0}=g =1.0$ and $T=2.0$.

\newpage
\begin{figure}
\centering
\includegraphics[width=390 pt]{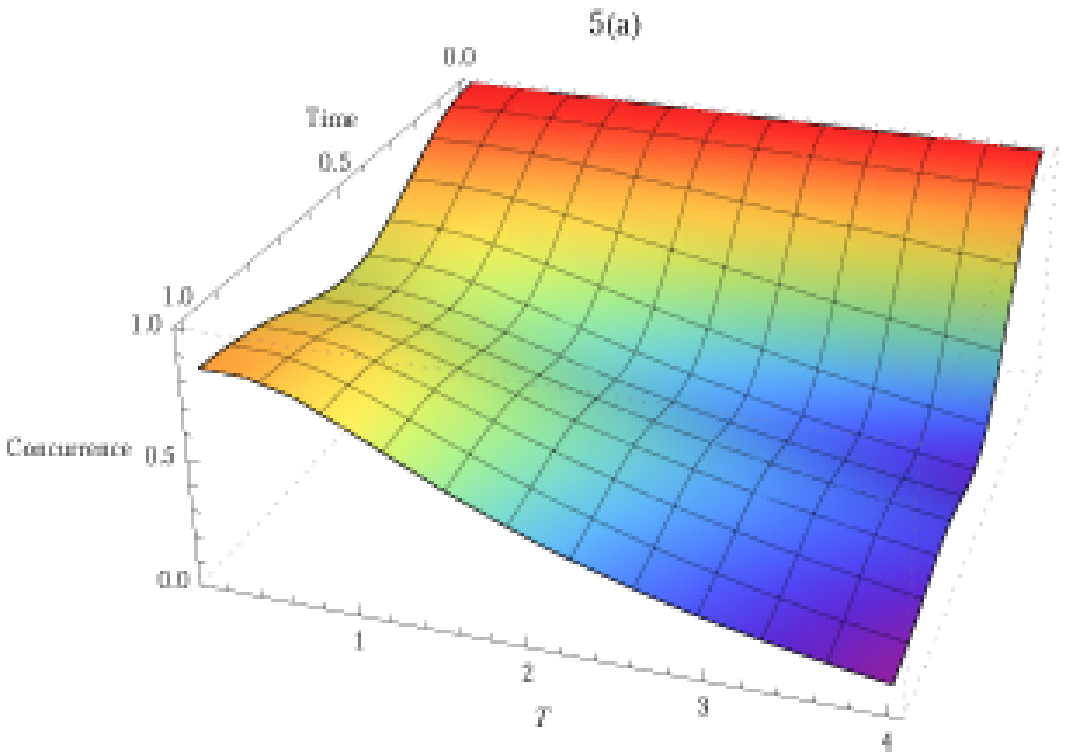}
\caption{}
\label{FIG. 5(a)}
\end{figure}
\textbf{Figure Caption}
\item FIG. 5(a):(Color online) The time evolution of the entanglement versus T for initial two-qubit state $|\psi_{s}(0)\rangle=\frac{1}{\surd2}( |00\rangle+ |11\rangle)$  for the  antiferromagnetic case $J=2.0$ and $J_{z}=1.0$. Other parameters are $\varepsilon=0.5, \,g_{0}=g =1.0$ and $D_{z}=2.0$.

\newpage
\begin{figure}
\centering
\includegraphics[width=390 pt]{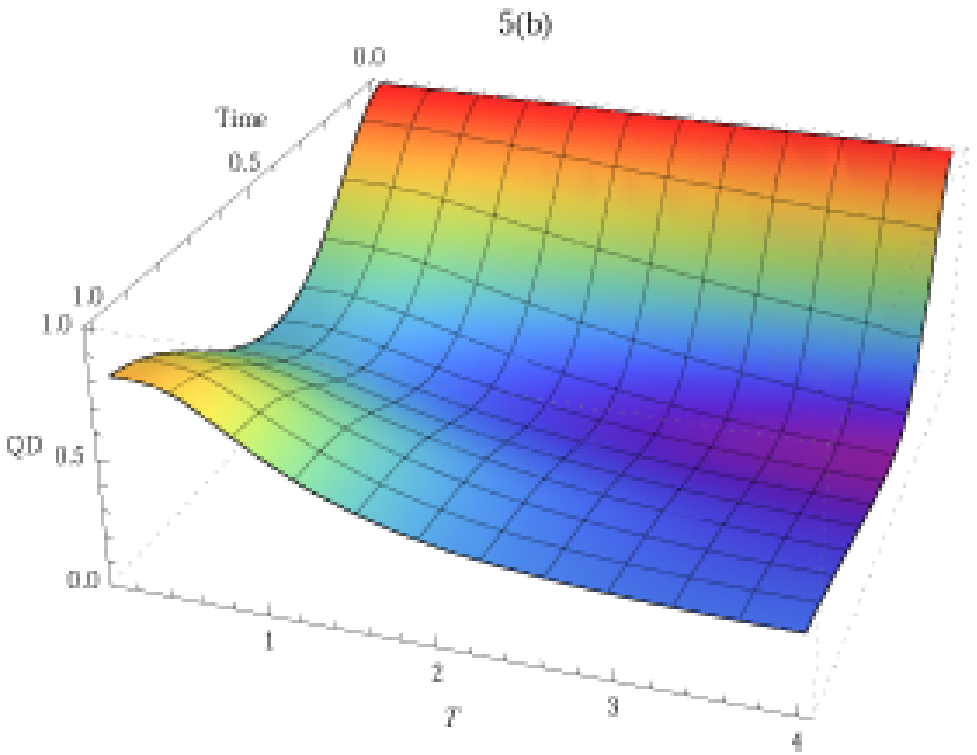}
\caption{}
\label{FIG. 5(b)}
\end{figure}
\textbf{Figure Caption}
\item Fig. 5(b):(Color online) The time evolution of the quantum discord versus T for initial two-qubit state $|\psi_{s}(0)\rangle=\frac{1}{\surd2}( |00\rangle+ |11\rangle)$  for the  antiferromagnetic case $J=2.0$ and $J_{z}=1.0$. Other parameters are $\varepsilon=0.5, \,g_{0}=g =1.0$ and $D_{z}=2.0$.

\end{document}